\documentclass[twocolumn,english,pra]{revtex4}
\usepackage[T1]{fontenc}
\usepackage[latin9]{inputenc}
\usepackage{amsmath}
\usepackage{graphicx}
\usepackage{amssymb}
\usepackage{esint}

\makeatletter
\@ifundefined{textcolor}{}
{%
 \definecolor{BLACK}{gray}{0}
 \definecolor{WHITE}{gray}{1}
 \definecolor{RED}{rgb}{1,0,0}
 \definecolor{GREEN}{rgb}{0,1,0}
 \definecolor{BLUE}{rgb}{0,0,1}
 \definecolor{CYAN}{cmyk}{1,0,0,0}
 \definecolor{MAGENTA}{cmyk}{0,1,0,0}
 \definecolor{YELLOW}{cmyk}{0,0,1,0}
 }

\makeatother

\makeatother

\usepackage{babel}

\makeatother

\usepackage{babel}

\begin{document}

\title{Confinement-induced resonance in quasi-one-dimensional systems under
transversely anisotropic confinement}

\author{Shi-Guo Peng$^{1,2}$, Seyyed S. Bohloul$^{2}$, Xia-Ji Liu$^{2}$,
Hui Hu$^{2}$, and Peter D. Drummond$^{2}$}

\affiliation{$^{1}$\ Department of Physics, Tsinghua University, Beijing 100084,
China\\
 $^{2}$ARC Centre of Excellence for Quantum-Atom Optics, Centre
for Atom Optics and Ultrafast Spectroscopy, Swinburne University of
Technology, Melbourne 3122, Australia }

\date{\today}
\begin{abstract}
We theoretically investigate the confinement-induced resonance for
quasi-one-dimensional quantum systems under transversely anisotropic
confinement, using a two-body \emph{s}-wave scattering model in the
zero-energy collision limit. We predict a single resonance for any
transverse anisotropy, whose position shows a slight downshift with
increasing anisotropy. We compare our prediction with the recent experimental
result by Haller \textit{et al.} {[}Phys. Rev. Lett. \textbf{104},
153203 (2010){]}, in which two resonances are observed in the presence
of transverse anisotropy. The discrepancy between theory and experiment
remains to be resolved.
\end{abstract}
\maketitle

\section{Introduction}

Recently ultracold low-dimensional atomic gases have attracted a great
deal of interest as they show unique quantum signatures not encountered
in three dimension (3D) \cite{LowDimRMP}. For example, a one-dimensional
(1D) gas of bosonic atoms with strongly repulsive interparticle interactions
acquires fermionic properties \cite{Petrov}. A 1D gas of fermionic
atoms may exhibit exotic inhomogeneous superfluidity \cite{hldprl2007,xiajifflo2007,xiajifflo2008}
and cluster-pairing \cite{xiajiclusterpairing}. In two dimensions,
the Berezinskii-Kosterlitz-Thouless (BKT) phase transition is very
different from the usual finite-temperature phase transitions \cite{Dalibard}.
Experimentally, ultracold low-dimensional atomic systems are conveniently
accessible upon introducing tight confinement via optical lattices
that remove one or two spatial degrees of freedom \cite{LowDimRMP}.
Furthermore, the interactions between atoms can also be tuned precisely
and arbitrarily by a collisional Feshbach resonance \cite{Feshbach1962,FRreview}.
Using these means, low-dimensional atomic systems now provide an ideal
experimental model to test fundamental many-body physics, and are
much simpler than corresponding systems in condensed matter physics.

Confinement-induced resonance (CIR) is an intriguing phenomenon found
when one relates the effective 1D coupling constant of quasi-1D quantum
gases, $g_{1D}$, to the 3D \emph{s}-wave scattering length, $a_{3D}$
\cite{Olshanii1998,Bergeman2003,Kim2005,Mora,Naidon2007,Saeidian}.
Considering a tight transverse confinement ($\omega_{\perp}=\omega_{x}=\omega_{y}$)
with the characteristic length scale $a_{\perp}=\sqrt{\hbar/(m\omega_{\perp})}$,
Olshanii showed that the effective 1D coupling strength diverges at
a particular ratio of $a_{3D}/a_{\perp}$ \cite{Olshanii1998}. The
underlying physics of the CIR is very similar to the well-known Feshbach
resonance \cite{Bergeman2003}, if we assume that under tight confinement
the ground-state transverse mode and the other transverse modes play
respectively the roles of the scattering {}``open'' channel and
{}``closed'' channels. The scattering state of two colliding atoms
in the ground-state transverse mode can be brought into resonance
with a bound molecular state in other high-lying transverse modes
by tuning the ratio of $a_{3D}/a_{\perp}$. This causes a confinement
induced resonance.

This type of novel resonance was recently confirmed in several ultracold
atom experiments \cite{Kinoshita,Paredes,Esslinger,Haller2009,Haller2010}.
In particular, the properties of CIR for an ultracold quantum gas
of Cs atoms were studied systematically by measuring atom loss and
heating rate under transversely anisotropic confinement near a rather
narrow Feshbach resonance \cite{Haller2010}. Upon increasing the
transverse anisotropy, two CIRs were observed, splitting from the
single CIR in the isotropic limit.

In this paper, we theoretically investigate the CIR under transversely
anisotropic confinement, using a zero-energy pseudopotential approach
\cite{Olshanii1998}, which is a useful generalization of Olshanii's
method. We find a single resonance in the scattering amplitude, except
with a downshift of the resonance position with increasing transverse
anisotropy. We calculate also the bound state of the decoupled excited
transverse manifold and find only one such bound state exists. Taken
together, we believe that there is only one CIR in an anisotropic
two-body \emph{s}-wave scattering model at zero-energy limit. The
two (splitting) resonances found experimentally therefore must be
caused by some other reasons beyond the simple \textit{s}-wave, zero-energy
scattering model.

In the following, we first present the essence of \textit{s}-wave
scattering approach and calculate the bound state of the excited transverse
manifold (Sec. II) and then discuss in detail the CIR as a function
of the transverse anisotropy (Sec. III). Sec. IV is devoted to conclusions
and further remarks.

\section{CIR under transversely anisotropic confinement}

\subsection{\label{sub:The-scattering-amplitude}The scattering amplitude}

In this section, we calculate the effective 1D coupling constant under
transversely anisotropic confinement by generalizing the zero-energy
\textit{s}-wave scattering approach developed by Olshanii \cite{Olshanii1998}.
Let us consider two atoms in a harmonic trap with weak confinement
in $z$-axis and tight confinement in the transverse direction such
that $\omega_{x}=\eta\omega_{y}\gg\omega_{z}$. The transverse anisotropy
is characterized by the ratio $\eta=\omega_{x}/\omega_{y}$ . The
motion along the $z$-axis is approximately free. To consider the
scattering in this direction, we therefore set $\omega_{z}=0$. Owing
to the separability of center-of-mass motion and relative motion in
harmonic traps, the low-energy \textit{s}-scattering along z-axis
is described by the following single-channel Hamiltonian: \begin{equation}
\hat{H}_{rel}=-\frac{\hbar^{2}}{2\mu}\frac{\partial^{2}}{\partial z^{2}}+{\cal \hat{H}}_{\bot}+g_{3D}\delta({\bf r})\frac{\partial}{\partial r}r,\label{eq:3D relative Hamiltonian}\end{equation}
 where \begin{equation}
\hat{H}_{\perp}=-\frac{\hbar^{2}}{2\mu}\left(\frac{\partial^{2}}{\partial x^{2}}+\frac{\partial^{2}}{\partial y^{2}}\right)+\frac{1}{2}\mu\omega_{y}^{2}\left(\eta^{2}x^{2}+y^{2}\right),\label{eq:3D x-y Hamiltonian}\end{equation}
 Here, $g_{3D}=4\pi\hbar^{2}a_{3D}/m$ is the 3D coupling constant,
and $\mu=m/2$ is the reduced mass for the relative motion. The effective
1D coupling constant $g_{1D}$ is obtained by matching the scattering
amplitude of the exact 3D solution $\Psi({\bf r})$ of Eq. (\ref{eq:3D relative Hamiltonian})
to that of a 1D Hamiltonian \begin{equation}
{\cal H}_{1D}=-\frac{\hbar^{2}}{2\mu}\frac{\partial^{2}}{\partial z^{2}}+g_{1D}\delta(z).\label{eq:1D Hamiltonian}\end{equation}
 The latter scattering amplitude $f_{e}(k_{z})$, with low scattering
energy $E=\hbar^{2}k_{z}^{2}/2\mu$, is then given by, \begin{equation}
\Psi(x=0,y=0,z)\propto e^{ik_{z}z}+f_{e}(k_{z})e^{ik_{z}\left|z\right|}\label{eq:asymptotic wavefunction}\end{equation}
 for $\left|z\right|\rightarrow\infty$ , and the effective 1D coupling
constant $g_{1D}$ can be determined by using \begin{equation}
g_{1D}=\lim_{k_{z}\rightarrow0}\frac{\hbar^{2}k_{z}}{\mu}\frac{\mathop{\rm Re}f_{e}(k_{z})}{\mathop{\rm Im}f_{e}(k_{z})}.\label{eq:definition of g1D}\end{equation}

Next, we shall calculate the 3D exact solution $\Psi({\bf r})$, from
which we can extract the effective 1D scattering amplitude $f_{e}(k_{z})$
at the collision energy $E=\hbar^{2}k_{z}^{2}/2\mu+\hbar\omega_{x}/2+\hbar\omega_{y}/2$,
when $k_{z}\rightarrow0$. Using the complete set of eigenstates of
the non-interacting Hamiltonian, the (un-normalized) scattering wavefunction
can be expanded as, \begin{equation}
\Psi({\bf r})=\int_{-\infty}^{\infty}dk\sum_{n_{1},n_{2}}A_{n_{1}n_{2}}\left(k\right)\phi_{n_{1}}(\frac{\sqrt{\eta}x}{d})\phi_{n_{2}}(\frac{y}{d})e^{ikz},\label{eq:scattering wavefunction}\end{equation}
 where $\phi_{n}(\xi)=(\sqrt{\pi}2^{n}n!)^{-1/2}e^{-\xi^{2}/2}H_{n}\left(\xi\right)$
is the $n$-th eigenstate of a 1D harmonic oscillator ($n=0,2,4,...$),
and $d=\sqrt{\hbar/\left(\mu\omega_{y}\right)}$ is the oscillator
length in the $y$-axis. In the summation, the integers $n_{1},n_{2}$
can be restricted to even numbers, as $s$-wave scattering preserves
parity. The expansion coefficients $A_{n_{1}n_{2}}(k)$ may be determined
by substituting the wavefunction (\ref{eq:scattering wavefunction})
into the Schrödinger equation $\hat{H}_{rel}\Psi({\bf r})=E\Psi({\bf r})$
and then projecting both sides of the equation onto noninteracting
states (i.e., the expansion basis) \cite{Busch,Idziaszek,Liang}.
Owing to the contact interaction, the expansion coefficient takes
the form,

\begin{equation}
A_{n_{1}n_{2}}(k)=-\frac{g_{3D}\phi_{n_{1}}^{*}(0)\phi_{n_{2}}^{*}(0){\cal F}}{n_{1}\hbar\omega_{x}+n_{2}\hbar\omega_{y}+\frac{\hbar^{2}}{2\mu}\left(k^{2}-k_{z}^{2}\right)},\label{eq:expansion coefficient}\end{equation}
 depending on a single parameter \begin{equation}
{\cal F}=\left[\frac{\partial\left(r\Psi({\bf r})\right)}{\partial r}\right]_{r=0}=\left[\frac{\partial\left(z\Psi(0,0,z)\right)}{\partial z}\right]_{z\rightarrow0^{+}}.\label{eq:BP condition}\end{equation}
 The summation over $n_{1},n_{2}$ and $k$ in the scattering wavefunction
$\Psi({\bf r})$ can then be performed \cite{Idziaszek,Liang}, yielding
\begin{eqnarray}
\frac{\Psi\left(0,0,z\right)}{\phi_{0}^{2}\left(0\right)} & = & e^{ik_{z}z}-i\frac{\eta^{1/4}\mu g_{3D}{\cal F}}{\sqrt{\pi}\hbar^{2}dk_{z}}e^{ik_{z}|z|}\nonumber \\
 &  & -\frac{\eta^{1/4}g_{3D}\mu{\cal F}}{2\sqrt{\pi}\hbar^{2}}\Lambda\left[\frac{2\left|z\right|}{d},-\left(\frac{dk_{z}}{2}\right)^{2}\right],\label{eq:scattering wavefunction 2}\end{eqnarray}
 where the function $\Lambda[\xi,\epsilon]$ takes the form \begin{equation}
\Lambda=\int\limits _{0}^{\infty}dt\frac{e^{-\epsilon t-\frac{\xi^{2}}{4t}}}{\sqrt{\pi t}}\left[\frac{1}{\sqrt{\left(1-e^{-\eta t}\right)\left(1-e^{-t}\right)}}-1\right].\end{equation}
 The parameter ${\cal F}$ is to be determined and it should satisfy
the boundary condition, Eq. (\ref{eq:BP condition}). We thus have
to check the asymptotic behavior of the wavefunction (\ref{eq:scattering wavefunction 2})
as $z\rightarrow0^{+}$, or to check $\Lambda[\xi,\epsilon]$ as $\xi\rightarrow0$.
After some algebra, one finds that, \begin{equation}
\Lambda[\xi,\epsilon]\stackrel{\xi\rightarrow0}{=}\frac{1}{\sqrt{\eta}}\left[\frac{2}{\xi}+{\cal L}\left(\epsilon\right)+{\cal L}_{1}\left(\epsilon\right)\xi+\cdots\right].\end{equation}
 At low energy, the zero-order term of ${\cal L}\left(\epsilon\right)$
has the form ${\cal L}\left(\epsilon\right)=-{\cal C}+\overline{{\cal L}}\left(\epsilon\right)$,
where the constant ${\cal C}$ is defined as \begin{equation}
{\cal C}=-\frac{1}{\sqrt{\pi}}\int\limits _{0}^{\infty}dt\left\{ \frac{\sqrt{\eta}}{\sqrt{t}}\left[\frac{1}{\sqrt{\left(1-e^{-\eta t}\right)\left(1-e^{-t}\right)}}-1\right]-\frac{1}{t^{3/2}}\right\} \label{eq:constant C}\end{equation}
and \begin{eqnarray}
\overline{{\cal L}}\left(\epsilon\right) & = & \sqrt{\frac{\eta}{\pi}}\sum_{n=1}^{\infty}\frac{\left(-\right)^{n}\epsilon^{n}}{n!}\times\nonumber \\
 &  & \int\limits _{0}^{\infty}dt\left[\frac{1}{\sqrt{\left(1-e^{-\eta t}\right)\left(1-e^{-t}\right)}}-1\right]t^{n-1/2}\label{eq:constant L}\end{eqnarray}
 Using the above asymptotic form of $\Lambda[\xi,\epsilon]$, the
parameter ${\cal F}$ can be easily determined, and hence the effective
1D scattering amplitude by comparing Eq (\ref{eq:scattering wavefunction 2})
and Eq. (\ref{eq:asymptotic wavefunction}). We finally arrive at,

\begin{eqnarray}
\frac{1}{f_{e}\left(k_{z}\right)} & = & -1+ik_{z}\frac{d^{2}}{2\sqrt{\eta}a_{3D}}\left(1-\frac{{\cal C}a_{3D}}{d}\right)\nonumber \\
 &  & +i(dk_{z}/2\sqrt{\eta})\overline{{\cal L}}\left(-\left(dk_{z}/2\right)^{2}\right).\label{eq:scattering amplitude}\end{eqnarray}
In the limit of an isotropic transverse confinement ($\eta=1$), the
constant ${\cal C}$ is \begin{equation}
{\cal C}=-\frac{1}{\sqrt{\pi}}\int\limits _{0}^{\infty}dt\left[\frac{1}{\sqrt{t}}\frac{1}{e^{t}-1}-\frac{1}{t^{3/2}}\right]=-\zeta\left(\frac{1}{2}\right).\label{eq:symmetric C}\end{equation}
 The function $\overline{{\cal L}}\left(\epsilon\right)$ has the
form \begin{equation}
\overline{{\cal L}}\left(\epsilon\right)=\frac{1}{\sqrt{\pi}}\sum_{n=1}^{\infty}\frac{\left(-\right)^{n}\epsilon^{n}}{n!}\int\limits _{0}^{\infty}dt\frac{t^{n-1/2}}{e^{t}-1}.\end{equation}
 Using the integral form of the Riemann zeta function \begin{equation}
\zeta\left(s\right)=\frac{1}{\Gamma\left(s\right)}\int\limits _{0}^{\infty}dx\frac{x^{s-1}}{e^{x}-1},\end{equation}
 we find that \begin{equation}
\overline{{\cal L}}\left(\epsilon\right)=\sum_{n=1}^{\infty}\left(-\right)^{n}\frac{\left(2n-1\right)!!}{2^{n}n!}\zeta\left(n+\frac{1}{2}\right)\epsilon^{n}.\label{eq:symmetric L}\end{equation}
 Eqs. (\ref{eq:symmetric C}) and (\ref{eq:symmetric L}) reproduce
the well-known result by Olshanii \cite{Olshanii1998}.

We can easily obtain the 1D coupling constant $g_{1D}$ from Eq.(\ref{eq:definition of g1D})
and (\ref{eq:scattering amplitude}), \begin{equation}
g_{1D}=\frac{2\hbar^{2}a_{3D}}{\mu d^{2}}\frac{\sqrt{\eta}}{1-{\cal C}(a_{3D}/d)},\end{equation}
 which diverges at $(a_{3D}^{(R)}/a_{y})=\sqrt{2}/{\cal C}$ , where
$a_{y}=\sqrt{\hbar/m\omega_{y}}=d/\sqrt{2}$ . Therefore, there is
only one CIR, whose position depends on the ratio $\eta=\omega_{x}/\omega_{y}$
for a given length scale $a_{y}$. At $\eta=1$, the constant in Eq.
(\ref{eq:constant C}) can be calculated analytically, giving ${\cal C}_{\eta=1}=-\zeta(1/2)=1.4603...$.
This recovers the well-known result given by Olshanii \cite{Olshanii1998}.

\subsection{Bound state of the excited transverse manifold}

\begin{figure}
\includegraphics[width=0.5\textwidth]{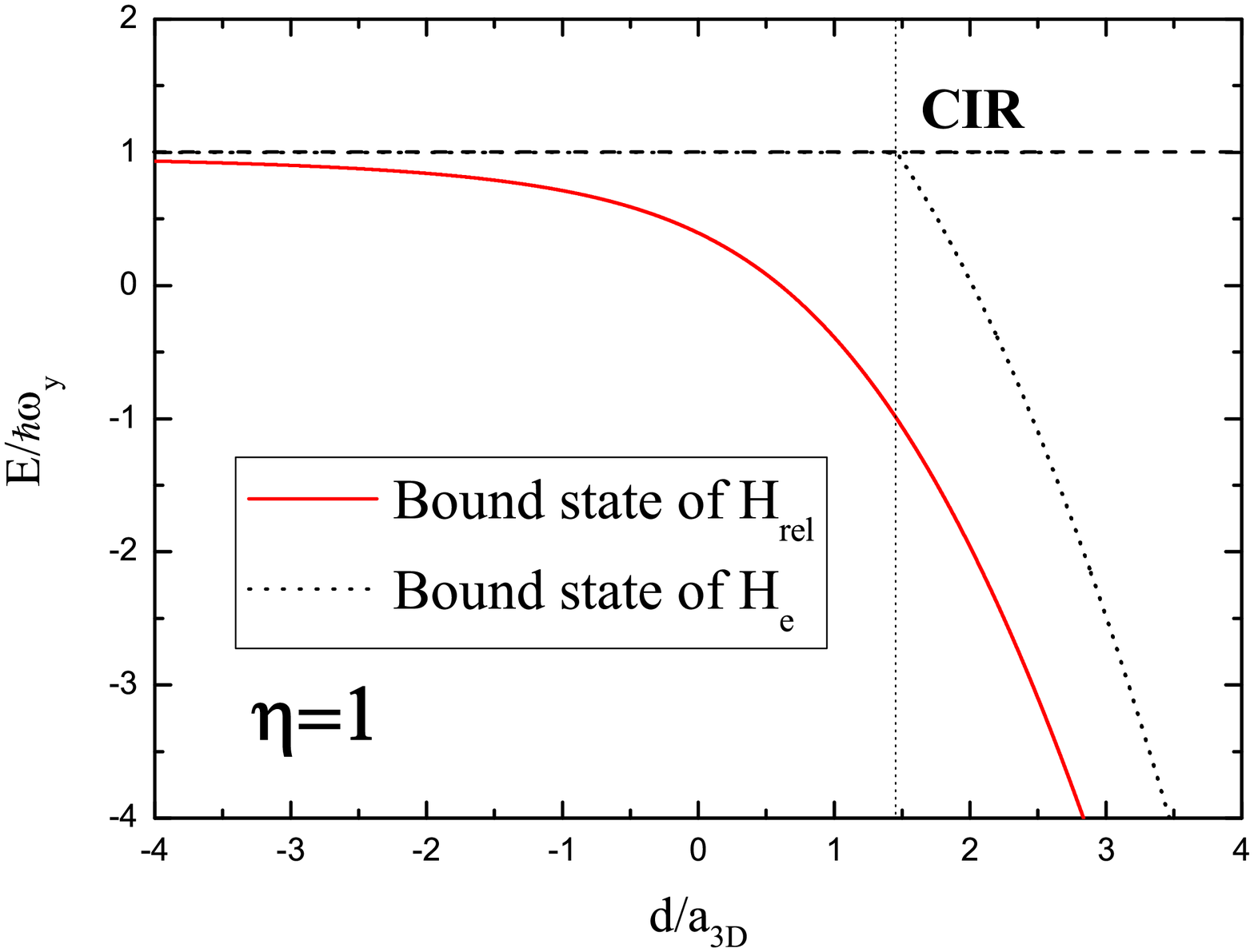}

\caption{(Color online) The bound states of the total relative Hamiltonian
$\hat{H}_{rel}$ (the solid curve) and the excited Hamiltonian $\hat{H}_{e}$
(the dotted curve) with $\eta=1$ .}

\label{Flo:fig1}
\end{figure}

\begin{figure}
\includegraphics[width=0.5\textwidth]{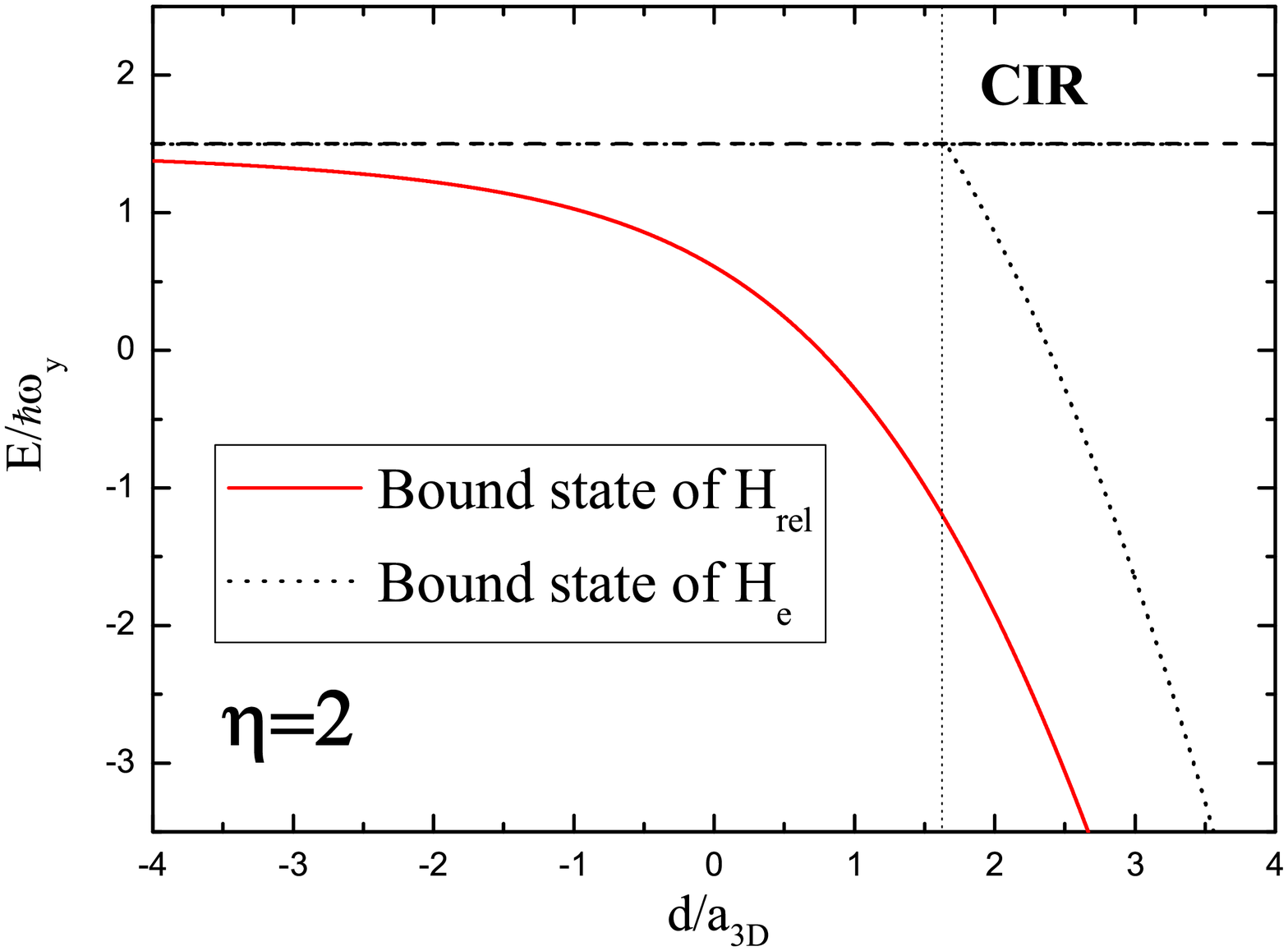}

\caption{(Color online) The same as Fig.\ref{Flo:fig1} but for $\eta=2$ .}

\label{Flo:fig2}
\end{figure}

Bergeman, \emph{et al. }\cite{Bergeman2003} show that the physical
origin of the CIR is a zero-energy Feshbach resonance under tight
transverse confinement. The ground transverse mode and the manifold
of excited modes play the roles of the open channel and the closed
channel, respectively. If a bound state exists in the closed channel,
a resonance will occur when the bound-state energy is equal to the
incident energy of two colliding atoms in the ground transverse mode
of the total Hamiltonian. In this section, we calculate the bound
state of the decoupled excited manifold. We find that a resonace will
appear at $a_{3D}^{(R)}/a_{y}=\sqrt{2}/{\cal C}$, consistent with
the calculations in Sec.\ref{sub:The-scattering-amplitude}.

Using a similar method to Sec.\ref{sub:The-scattering-amplitude},
the wavefunction of the total relative Hamiltonian $\hat{H}_{rel}$
can be expanded in terms of the eigenstates of the non-interacting
Hamiltonian, Eq.(\ref{eq:scattering wavefunction}). After substituting
Eq.(\ref{eq:scattering wavefunction}) into the Schroedinger equation
and using the normalization and orthogonality of the wavefunctions,
we arrive at,\begin{equation}
\left[\frac{\partial}{\partial r}r\mathcal{H}\left(\vec{r}\right)\right]_{r=0}=-\frac{1}{2\pi a_{3D}d^{2}}\label{eq:App.B.1}\end{equation}
 where\begin{equation}
\mathcal{H}\left(\vec{r}\right)=\int_{-\infty}^{\infty}\frac{dk_{z}}{2\pi}\sum_{n_{1}=0}^{\infty}\sum_{n_{2}=0}^{\infty}\frac{e^{ik_{z}z}\phi_{n_{1}}^{*}\left(0\right)\phi_{n_{2}}^{*}\left(0\right)\phi_{n_{1}}\left(x\right)\phi_{n_{2}}\left(y\right)}{\frac{1}{2}d^{2}k_{z}^{2}+n_{1}\eta+n_{2}-\epsilon}\label{eq:App.B.2}\end{equation}
and $\epsilon=E/\hbar\omega_{y}-\eta/2-1/2$ . Using \begin{equation}
\frac{1}{n}=\int_{0}^{1}dqq^{n-1}\qquad\left(n>0\right)\label{eq:App.B.3}\end{equation}
 and\begin{equation}
\sum_{k=0}^{\infty}\frac{t^{k}}{2^{k}k!}H_{k}(x)H_{k}(y)=\frac{exp(\frac{2txy-t^{2}x^{2}-t^{2}y^{2}}{1-t^{2}})}{\sqrt{1-t^{2}}}\,,\label{eq:App.B.4}\end{equation}
Eq.(\ref{eq:App.B.2}) can be written as\begin{eqnarray}
\mathcal{H}\left(\vec{r}\right) & = & \frac{\eta^{1/2}}{2\pi^{3/2}d^{3}}exp\left[-\frac{1}{2d^{2}}\left(\eta x^{2}+y^{2}\right)\right]\times\nonumber \\
 &  & \int_{0}^{\infty}dt\frac{exp\left[\frac{t\epsilon}{2}-\frac{1}{t}\frac{z^{2}}{d^{2}}\right]}{\sqrt{t}}\frac{exp\left[-\frac{\eta e^{-\eta t}x^{2}}{\left(1-e^{-\eta t}\right)d^{2}}\right]}{\sqrt{1-e^{-\eta t}}}\times\nonumber \\
 &  & \frac{exp\left[-\frac{e^{-t}y^{2}}{\left(1-e^{-t}\right)d^{2}}\right]}{\sqrt{1-e^{-t}}}\label{eq:App.B.5}\end{eqnarray}
 for $\epsilon<0$ . As $r\rightarrow0$ , only the small $t$ terms
dominate the integral, so that\begin{eqnarray}
\mathcal{H}\left(r\rightarrow0\right) & = & \frac{\eta^{1/2}}{2\pi^{3/2}d^{3}}\int_{0}^{\infty}dt\frac{exp\left[-\frac{r^{2}}{td^{2}}\right]}{\eta^{1/2}t^{3/2}}\nonumber \\
 & = & \frac{1}{2\pi d^{2}r}\,.\label{eq:App.B.6}\end{eqnarray}
 Hence, the function $\mathcal{H}\left(\vec{r}\right)$ can be rewritten
as,\begin{equation}
\mathcal{H}\left(\vec{r}\right)=\frac{1}{2\pi^{3/2}d^{3}}\mathcal{F}\left(\epsilon,\vec{r}\right)+\frac{1}{2\pi d^{2}r}\label{eq:App.B.7}\end{equation}
and the regular part $\mathcal{F}\left(\epsilon,\vec{r}\right)$ has
the form,\begin{eqnarray}
\mathcal{F}\left(\epsilon,\vec{r}\right) & = & exp\left[-\frac{1}{2d^{2}}\left(\eta x^{2}+y^{2}\right)\right]\times\nonumber \\
 &  & \int_{0}^{\infty}dt\left\{ \frac{\sqrt{\eta}exp\left[\frac{t\epsilon}{2}-\frac{1}{t}\frac{z^{2}}{d^{2}}\right]}{\sqrt{t}}\frac{exp\left[-\frac{\eta e^{-\eta t}x^{2}}{\left(1-e^{-\eta t}\right)d^{2}}\right]}{\sqrt{1-e^{-\eta t}}}\times\right.\nonumber \\
 &  & \frac{exp\left[-\frac{e^{-t}y^{2}}{\left(1-e^{-t}\right)d^{2}}\right]}{\sqrt{1-e^{-t}}}-\left.\frac{exp\left[-\frac{r^{2}}{td^{2}}\right]}{t^{3/2}}\right\} \,.\label{eq:App.B.8}\end{eqnarray}
Next, from Eq.(\ref{eq:App.B.1}), we obtain, \begin{equation}
\mathcal{F}\left(\epsilon,0\right)=-\frac{\sqrt{\pi}d}{a_{3D}}\label{eq:App.B.9}\end{equation}
 where\begin{equation}
\mathcal{F}\left(\epsilon,0\right)=\int_{0}^{\infty}dt\left\{ \frac{\sqrt{\eta}exp\left(\frac{t\epsilon}{2}\right)}{\sqrt{t\left(1-e^{-\eta t}\right)\left(1-e^{-t}\right)}}-\frac{1}{t^{3/2}}\right\} \label{eq:App.B.10}\end{equation}
 for $\epsilon<0$ . Eq.(\ref{eq:App.B.9}) and (\ref{eq:App.B.10})
determine the bound-state energy of the total relative Hamiltonian
depending on the interaction strength, which is denoted by the 3D
scattering length $a_{3D}$ .

If the transverse ground state is projected out from the Hilbert space,
we can define the excited Hamiltonian as,\begin{equation}
\hat{H}_{e}\equiv\hat{H}_{rel}-\hat{P_{g}}\hat{H}_{rel}\hat{P}_{g}\label{eq:2.B.3}\end{equation}
 where $\hat{P}_{g}=\left|g\right\rangle \left\langle g\right|$ is
the projection operator of the transverse ground mode $\left|g\right\rangle $
. We can also calculate the bound-state energy of the excited Hamiltonian
$\hat{H}_{e}$ by removing the $n_{1}=n_{2}=0$ term from the summation
of the function $\mathcal{H}\left(\vec{r}\right)$ , \emph{i.e.},
Eq.(\ref{eq:App.B.2}). Then the bound-state energy of the excited
Hamiltonian $\hat{H}_{e}$ should be determined by,\begin{equation}
\left[\frac{\partial}{\partial r}r\mathcal{H}_{e}\left(\vec{r}\right)\right]_{r=0}=-\frac{1}{2\pi a_{3D}d^{2}}\label{eq:App.B.11}\end{equation}
where\begin{eqnarray}
\mathcal{H}_{e}\left(\vec{r}\right) & = & \mathcal{H}\left(\vec{r}\right)-\frac{L\sqrt{\eta}}{\sqrt{2}\pi d^{3}\sqrt{-\epsilon}}exp\left[-\frac{1}{2d^{2}}\left(\eta x^{2}+y^{2}\right)\right]\times\nonumber \\
 &  & exp\left(-\sqrt{-2\epsilon}\frac{\left|z\right|}{d}\right)\label{eq:App.B.12}\end{eqnarray}
 for $\epsilon<0$ . Consequently, we can redefine the regular part
of $\mathcal{H}_{e}\left(\vec{r}\right)$ as\begin{eqnarray}
\mathcal{F}_{e}\left(\epsilon,\vec{r}\right) & = & \mathcal{F}\left(\epsilon,\vec{r}\right)-\sqrt{\frac{2\pi\eta}{-\epsilon}}exp\left[-\frac{1}{2d^{2}}\left(\eta x^{2}+y^{2}\right)\right]\times\nonumber \\
 &  & exp\left(-\sqrt{-2\epsilon}\frac{\left|z\right|}{d}\right)\,.\label{eq:App.B.13}\end{eqnarray}
 Then, using Eq.(\ref{eq:App.B.11}) and (\ref{eq:App.B.13}), we
arrive at\begin{equation}
\mathcal{F}_{e}\left(\epsilon,0\right)=-\frac{\sqrt{\pi}d}{a_{3D}}\label{eq:App.B.14}\end{equation}
 where\begin{eqnarray*}
\mathcal{F}_{e}\left(\epsilon,0\right) & = & \int_{0}^{\infty}dt\left\{ \frac{exp\left(\frac{t\epsilon}{2}\right)\sqrt{\eta}}{\sqrt{t}}\times\right.\\
 &  & \left.\left[\frac{1}{\sqrt{\left(1-e^{-\eta t}\right)\left(1-e^{-t}\right)}}-1\right]-\frac{1}{t^{3/2}}\right\} \,.\end{eqnarray*}
This determines the bound-state energy of the excited Hamiltonian
$\hat{H}_{e}$ .

For isotropic confinement ($\eta=1$), it is obvious that,\begin{equation}
\mathcal{F}_{e}\left(\epsilon,0\right)=\mathcal{F}\left(\epsilon-2\hbar\omega,0\right)\end{equation}
 Hence, from Eq.(\ref{eq:App.B.9}) and (\ref{eq:App.B.14}), the
bound state of the excited Hamiltonian $\hat{H}_{e}$ is $2\hbar\omega$
higher than that of the total relative Hamiltonian $\hat{H}_{rel}$
, which is consistent with the conclusion of Bergeman \cite{Bergeman2003}.

The bound states of the total relative Hamiltonian $\hat{H}_{rel}$
and the excited Hamiltonian $\hat{H}_{e}$ are plotted as functions
of the 3D scattering length $a_{3D}$ in Fig.\ref{Flo:fig1} and Fig.\ref{Flo:fig2}
with transverse aspect ratios $\eta=1$ and $\eta=2$ for isotropic
and the anisotropic confinement, respectively. It is obvous that there
is only one bound state for the excited Hamiltonian $\hat{H}_{e}$
both for isotropic and anisotropic confinement. A confinement-induced
resonance occurs when this bound state becomes degenerate with the
threshold of the transverse ground mode of the total relative Hamiltonian.
The position of the CIR in terms of the 3D scattering length is determined
by Eq.(\ref{eq:App.B.14}) with $\epsilon=0$,\begin{eqnarray}
\frac{d}{a_{3D}^{(R)}} & = & -\frac{1}{\sqrt{\pi}}\int_{0}^{\infty}dt\left\{ \frac{\sqrt{\eta}}{\sqrt{t}}\left[\frac{1}{\sqrt{\left(1-e^{-\eta t}\right)\left(1-e^{-t}\right)}}-1\right]\right.\nonumber \\
 &  & \left.-\frac{1}{t^{3/2}}\right\} \nonumber \\
 & \equiv & {\cal C}.\label{eq:2.B.6}\end{eqnarray}
 This result is consistent with that in the Sec.\ref{sub:The-scattering-amplitude}
which is $(a_{3D}^{(R)}/a_{y})=\sqrt{2}/{\cal C}$.

\section{Results and discussions}

\begin{figure}
\includegraphics[width=0.5\textwidth]{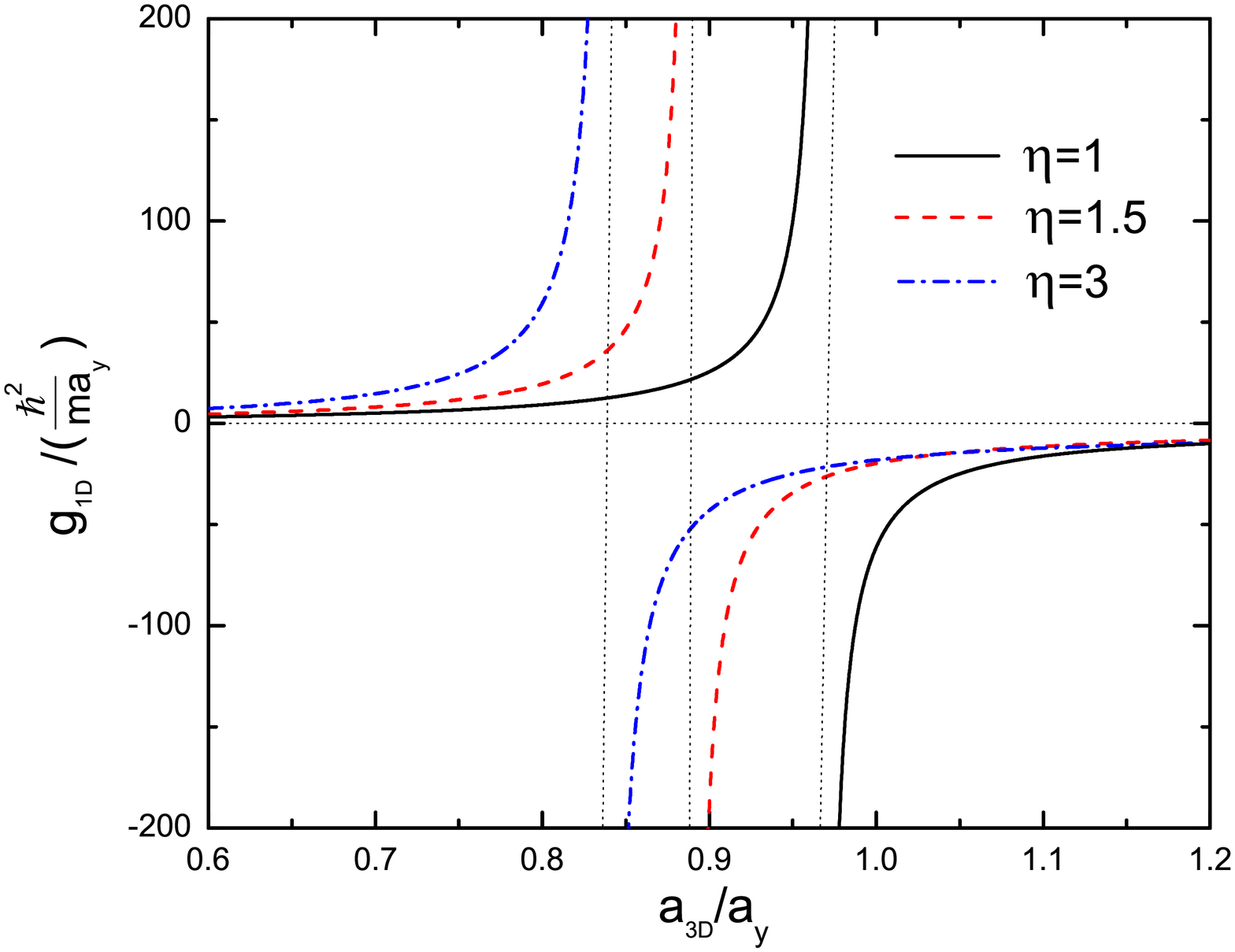}

\caption{(Color online) The effective 1D coupling constant, $g_{1D}$ (in units
of $\hbar^{2}/(ma_{y})$), as a function of $a_{3D}/a_{y}$ at different
transverse anisotropy $\eta=1$ (solid line), $1.5$ (dashed line),
and $3$ (dot-dashed line).}

\label{fig3} 
\end{figure}

We plot in Fig.\ref{fig3} the effective 1D coupling constant as a
function of the 3D \emph{s}-wave scattering length at $\eta=1$, $1.5$,
and $3$. In this parameter range, with increasing transverse anisotropy,
the resonance position of CIR shifts from $(a_{3D}/a_{y})_{\eta=1}^{(R)}=\sqrt{2}{\cal C}_{\eta=1}^{-1}=0.9684...$
to lower values. For the experimental situation with a 3D scattering
length tuned by a magnetic Feshbach resonance, this decrease corresponds
to a shift to the deep BEC side of the Feshbach resonance.

\begin{figure}
\includegraphics[width=0.5\textwidth]{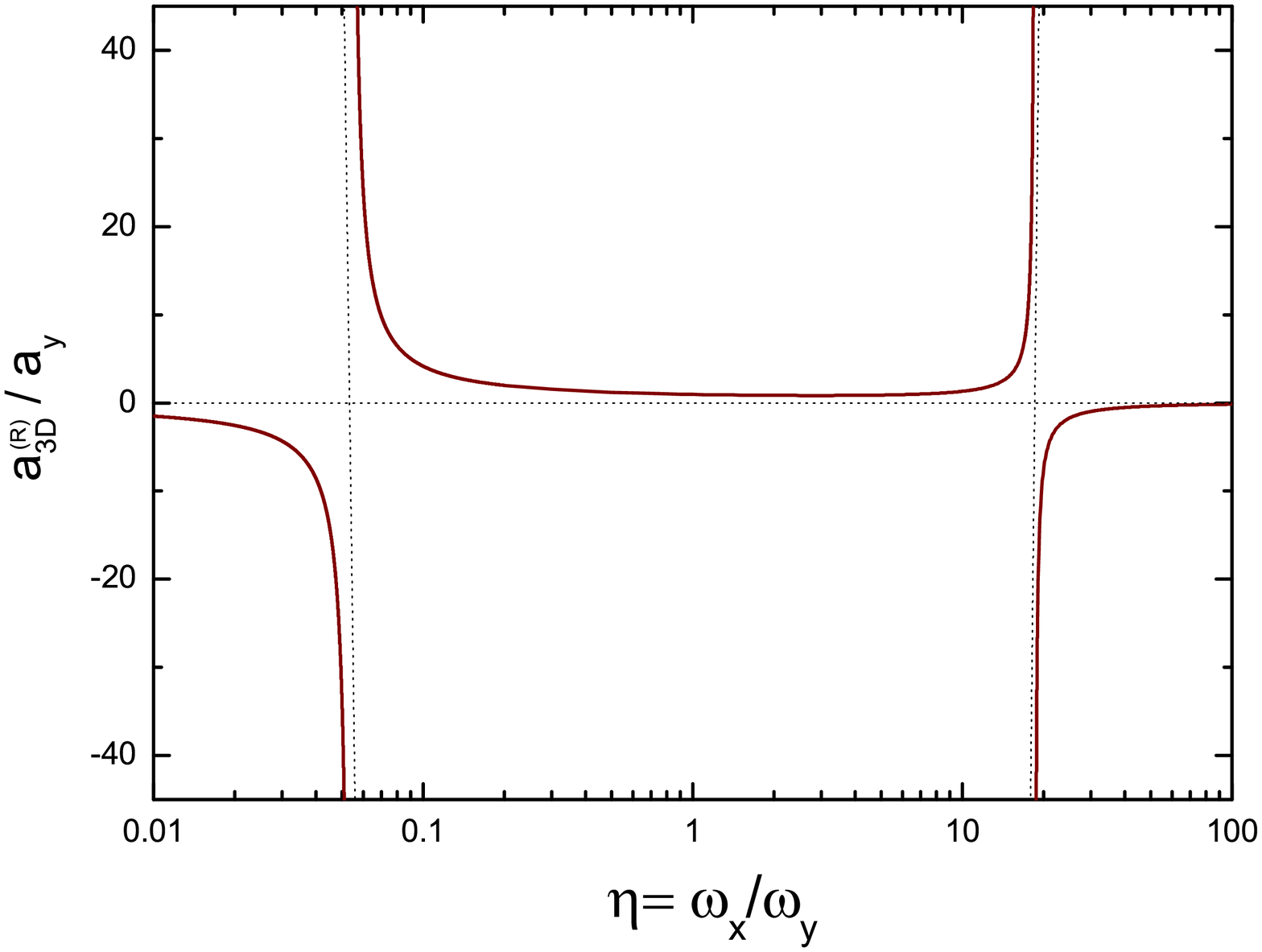}

\caption{(Color online) Resonance positions of CIR, $a_{3D}^{(R)}/a_{y}$,
as a function of the transverse anisotropy $\eta=\omega_{x}/\omega_{y}$.
For sufficient large or small values of transverse anisotropy, the
quasi-1D system crosses over to the quasi-2D regime.}

\label{fig4} 
\end{figure}

In a wider parameter space, however, we observe an interesting \emph{non-monotonic}
behavior of the resonance position as a function of the ratio $\eta$.
As shown in Fig.\ref{fig4}, at sufficiently large (or small) values
of the transverse anisotropy $\eta$, the resonance position $a_{3D}^{(R)}$
increases with increasing (or decreasing) transverse anisotropy, and
diverges at a critical frequency ratio $\eta_{0}\simeq18.551$ (or
$\eta_{0}^{-1}\simeq0.0539$). For $a_{3D}^{(R)}$, this corresponds
to a \emph{bend-back} to the magnetic Feshbach resonance from the
BEC limit. On further increasing (or decreasing) the transverse anisotropy,
the quasi-1D quantum gas crosses over to a pancake-shaped, horizontally-oriented
2D system, which would lead to a negative value for $a_{3D}^{(R)}$.
This observation is in agreement with theoretical calculations\cite{Naidon2007}
showing that for a quasi-2D system, the $s$-wave CIR occurs always
at a \emph{negative} 3D scattering length ($a_{3D}<0$) or on the
BCS side of the magnetic Feshbach resonance. 

\begin{figure}
\includegraphics[width=0.5\textwidth]{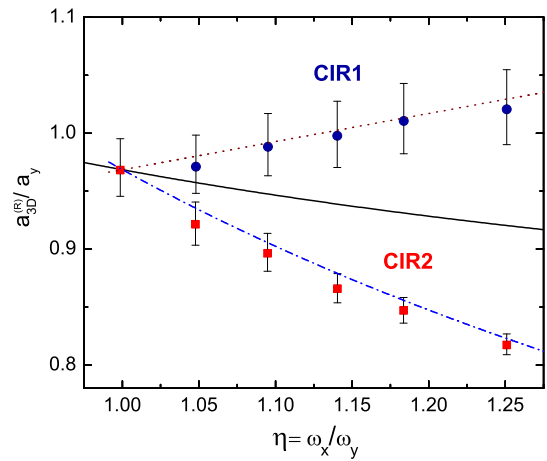}

\caption{(Color online) Resonance positions of CIRs, obtained by solving $a_{3D}^{(R)}/a_{y}=\sqrt{2}/\mathcal{C}$,
as a function of the transverse anisotropy $\eta=\omega_{x}/\omega_{y}$.
The experimental results (symbols) are compared with our zero-energy
\textit{s}-wave scattering prediction (solid line). In the presence
of transverse anisotropy, it is difficult to determine directly the
resonance positions from atom loss. Therefore, experimentally the
positions were obtained by determining the associated atom number
minima and then subtracting a constant offset. In accord with this
procedure, we have uniformly offset the experimental data in such
a way that the measured resonance position in the symmetric limit
($\eta=1$) is equal to the known theoretical prediction by Olshanii
\cite{Olshanii1998}. The dashed and dot-dashed lines are the resonance
positions, obtained respectively by solving, $a_{3D}^{(R)}/a_{y}=\sqrt{2\eta}/\mathcal{C}$
and $a_{3D}^{(R)}/a_{y}=\sqrt{2/\eta}/\mathcal{C}$. See the text
for more details. }

\label{Flo:fig1e} 
\end{figure}

We now compare our theoretical results with the experimental data
\cite{Haller2010}. This experiment uses a narrow Feshbach resonance
in $^{133}Cs$ \cite{Kraemer2004}, together with a nonadiabatic jump
to a particular magnetic field to initiate the resonance with a finite
kinetic energy. As a result, one may expect some differences to the
simplest single-channel model used here. The single-channel model
does not include the molecular field corresponding to the bound molecule
state. It is known from both mean-field \cite{Diener2004} and beyond
mean-field \cite{xiajiggtheory} calculations that, the single-channel
model describes only a broad Feshbach resonance. As illustrated in
Fig.\ref{Flo:fig1e}, our prediction of $a_{3D}^{(R)}$ is just between
the two CIRs observed in the experiment, and disagrees with either
of them. In fact, the observation of a single CIR \textit{s}-wave
resonance in our theoretical calculation can be easily understood.
Recall that the CIR is a kind of Feshbach resonance, which occurs
when the scattering state in an open channel becomes degenerate in
energy with the bound states in the closed channel. The number of
CIRs is therefore simply equal to the number of bound states in the
closed channel. For a quasi-1D quantum gas, the excited transverse
modes are grouped together and regarded as closed channels (see, for
example, ref. \cite{Bergeman2003} for a detailed discussion of closed
channels). In our model, the sub-Hilbert space of the excited transverse
modes can only sustain a \emph{single} bound state, no matter how
large large the transverse anisotropy is. The experimental observation
of two resonances must therefore have a different origin than the
simple \textit{\emph{zero-energy}}\textit{ s}-wave scattering picture
pursued in this work.

Interestingly, we obtain a good agreement with experimental data if
we include in the resonance condition an additional factor of $\sqrt{\eta}$.
As shown in Fig. 5, the CIR$_{1}$ and CIR$_{2}$ resonances are well
reproduced theoretically if we use the conditions, $a_{3D}^{(R)}/a_{y}=\sqrt{2\eta}/\mathcal{C}$
and $a_{3D}^{(R)}/a_{y}=\sqrt{2/\eta}/\mathcal{C}$, respectively.
This additional factor however can not justified. 

It should be note that, by keeping $\omega_{x}$ invariant and weakening
the confinement $\omega_{y}$ - which therefore gives an alternative
way to increase the value of $\eta$ - the observed CIR$_{1}$ displays
additional structure. It further splits into many sub-resonances at
$\eta>1.5$ and finally disappears at sufficiently large values of
transverse anisotropy. At the same time the CIR$_{2}$ persists. Also,
in the limit of a 2D system ($\eta\rightarrow\infty$), the resonance
position of the surviving CIR$_{2}$ is at a \emph{positive }3D scattering
length. These experimental observations also disagree with our simple
zero-energy \textit{s}-wave scattering calculation, and with previous
calculations\cite{Naidon2007} of 2D s-wave CIR. One possibility is
that many-body physics of the quantum bosonic gas may come into play.
A likely explanation therefore is that the CIRs observed in this experiment
are associated with the failure of Olshanii's model \cite{Olshanii1998}
which is only a zero-energy single-channel \textit{s}-wave description
of the inter-atomic interactions, and the experimental conditions
are different to this simple model. An advanced theory of CIRs need
to be introduced, \emph{e.g.}, multi-channel scattering theory including
molecular channels, non-zero collision energy theory, or many-body
effects.

\section{Conclusions and remarks}

In conclusion, we have presented a theoretical calculation of confinement-induced
resonance (CIR) under transversely anisotropic confinement, by extending
the standard zero-energy \textit{s}-wave scattering approach to incorporate
transverse anisotropy. We have theoretically found a single CIR and
have explained its physical origin. The position of our predicted
CIR disaggrees with either of the two CIRs measured most recently
in a quasi-1D quantum bosonic gas of $^{133}Cs$ atoms. Therefore,
the CIRs observed in the experiment cannot be explained simply by
the zero-energy \emph{s}-wave scattering theory. There are also several
discrepancies when the quasi-1D system crosses over to the quasi-2D
regime, which may arise either from the many-body physics of Cs atoms,
or the residual molecular fraction and angular momentum of the magnetic
Feshbach resonance. To take these aspects into account, we look forward
to improving the current \textit{s}-wave scattering calculation by
replacing the two-body scattering matrix with its many-body counterpart
and/or adopting a two-channel model to describe the magnetic Feshbach
resonance.
\begin{acknowledgments}
We gratefully acknowledge valuable discussions with Hui Dong. This
research was supported by the Australian Research Council (ARC) Center
of Excellence, ARC Discovery Project Nos. DP0984522 and DP0984637,
and the NFRP-China Grant (973 Project) Nos. 2006CB921404 and 2006CB921306.\end{acknowledgments}


\begin{thebibliography}{26}
\bibitem{LowDimRMP} I. Bloch, J. Dalibard, and W. Zwerger, Rev. Mod.
Phys. \textbf{80}, 885 (2008).

\bibitem{hldprl2007} H. Hu, X.-J. Liu, and P. D. Drummond, Phys.
Rev. Lett. \textbf{98}, 070403 (2007).

\bibitem{xiajifflo2007} X.-J. Liu, H. Hu, and P. D. Drummond, Phys.
Rev. A \textbf{76}, 043605 (2007).

\bibitem{xiajifflo2008} X.-J. Liu, H. Hu, and P. D. Drummond, Phys.
Rev. A \textbf{78}, 023601 (2008).

\bibitem{xiajiclusterpairing} X.-J. Liu, H. Hu, and P. D. Drummond,
Phys. Rev. A \textbf{77}, 013622 (2008).

\bibitem{Petrov} D. S. Petrov, G. V. Shlyapnikov, and J. T. M. Walraven,
Phys. Rev. Lett. \textbf{85}, 3745 (2000).

\bibitem{Dalibard} Z. Hadzibabic, P. Krüger, M. Cheneau, B. Battelier,
J. Dalibard, Nature \textbf{441}, 1118 (2006).

\bibitem{Feshbach1962} H. Feshbach, Ann. Phys. \textbf{19}, 287 (1962).

\bibitem{FRreview} T. Koehler, K. Goral, and P. S. Julienne, Rev.
Mod. Phys. \textbf{78}, 1311 (2006).

\bibitem{Olshanii1998} M. Olshanii, Phys. Rev. Lett. \textbf{81},
938 (1998).

\bibitem{Bergeman2003} T. Bergeman, M. G. Moore, and M. Olshanii,
Phys. Rev. Lett. \textbf{91,} 163201(2003).

\bibitem{Kim2005} J. I. Kim, J. Schmiedmayer, and P. Schmelcher,
Phys. Rev. A \textbf{72}, 042711(2005).

\bibitem{Mora} C. Mora, R. Egger, and A. O. Gogolin, Phys. Rev. A
\textbf{71}, 052705 (2005).

\bibitem{Naidon2007} P. Naidon, E. Tiesinga, W. F. Mitchell, and
P. S Julienne, New J. Phys. \textbf{9}, 19 (2007).

\bibitem{Saeidian} S. Saeidian, V. S. Melezhik, and P. Schmelcher,
Phys. Rev. A \textbf{77}, 042721 (2008).

\bibitem{Kinoshita} T. Kinoshita, T. Wenger, and D. S. Weiss, Science
\textbf{305}, 1125 (2004).

\bibitem{Paredes} B. Paredes, A. Widera, V. Murg, O. Mandel, S. Fölling,
I. Cirac, G. V. Shlyapnikov, T. W. Hänsch, and I. Bloch, Nature \textbf{429},
277 (2004).

\bibitem{Esslinger} K. Günter, T. Stöferle, H. Moritz, M. Köhl, and
T. Esslinger, Phys. Rev. Lett. \textbf{95}, 230401(2005).

\bibitem{Haller2009} E. Haller, M. Gustavsson, M. J. Mark, J. G.
Danzl, R. Hart, G. Pupillo, and H.-C. Nägerl, Science \textbf{325},
1224 (2009).

\bibitem{Haller2010} E. Haller, M. J. Mark, R. Hart, J. G. Danzl,
L. Reichsöllner, V. Melezhik, P. Schmelcher, and H.-C. Nägerl, Phys.
Rev. Lett. \textbf{104}, 153203(2010).

\bibitem{Busch} T. Busch, B.-G. Englert, K. Rza\.{z}ewski, and M.
Wilkens, Found. Phys. \textbf{28}, 549 (1998).

\bibitem{Idziaszek} Z. Idziaszek and T. Calarco, Phys. Rev. A \textbf{74},
022712 (2006).

\bibitem{Liang} J.-J. Liang and C. Zhang, Phys. Scr. \textbf{77},
025302 (2008).

\bibitem{Kraemer2004} T. Kraemer, J. Herbig, M. Mark, T. Weber, C.
Chin, H.-C. Nagerl and R. Grimm, App. Phys. B \textbf{79}, 1013 (2004).

\bibitem{Diener2004} R. B. Diener and T.-L. Ho, arXiv:0405174 (2004).

\bibitem{xiajiggtheory} X.-J. Liu and H. Hu, Phys. Rev. A \textbf{72},
063613 (2005).
\end{thebibliography}
\end{document}